# Superconductor Digital Electronics: Scalability and Energy Efficiency Issues

Sergey K. Tolpygo

*Abstract*—Superconductor digital electronics using Josephson junctions as ultrafast switches and magnetic-flux encoding of information was proposed over 30 years ago as a sub-terahertz clock frequency alternative to semiconductor electronics based on complementary metal-oxide-semiconductor (CMOS) transistors. Recently, interest in developing superconductor electronics has been renewed due to a search for energy saving solutions in applications related to high-performance computing. The current state of superconductor electronics and fabrication processes are reviewed in order to evaluate whether this electronics is scalable to a very large scale integration (VLSI) required to achieve computation complexities comparable to CMOS processors. A fully planarized process at MIT Lincoln Laboratory, perhaps the most advanced process developed so far for superconductor electronics, is used as an example. The process has nine superconducting layers: eight Nb wiring layers with the minimum feature size of 350 nm, and a thin superconducting layer for making compact high-kinetic-inductance bias inductors. All circuit layers are fully planarized using chemical mechanical planarization (CMP) of $SiO_2$ interlayer dielectric. The physical limitations imposed on the circuit density by Josephson junctions, circuit inductors, shunt and bias resistors, etc., are discussed. Energy dissipation in superconducting circuits is also reviewed in order to estimate whether this technology, which requires cryogenic refrigeration, can be energy efficient. Fabrication process development required for increasing the density of superconductor digital circuits by a factor of ten and achieving densities above $10^7$ Josephson junctions per $cm^2$ is described.

## I. INTRODUCTION

MAY 3, 2016 is the 100th anniversary of birth of Kirill Borisovich Tolpygo, a prominent theoretical physicist widely recognized for his contributions to condensed matter physics, crystal lattice dynamics, physics of semiconductors and dielectrics, and also biophysics [1]-[5]. Among his many works on application of mathematical and quantum-mechanical methods to biological systems, a significant part was devoted to developing an understanding of the mechanisms of high energy efficiency of living organisms, in particular the mechanisms of chemical energy conversion into mechanical energy in muscles and muscle contraction. In 1978 Tolpygo proposed a mechanism of muscle contraction that

Corresponding author: Sergey K. Tolpygo (e-mail: sergey.tolpygo@ll.mit.edu).

The author is with Lincoln Laboratory, Massachusetts Institute of Technology, Lexington, MA 02420, USA.

results from a sequential transfer of proton excitation, a proton exciton, along a chain of hydrogen bonds between two biopolymers, an actin-myosin pair [6],[7]. A pulling force is produced due to lowering the excited proton energy and shortening the bond length [8]. The initial excitation is provided by hydrolysis of an adenosine-triphosphate (ATP) molecule. A high energy efficiency of muscle contraction is explained in this model as due to energy recycling - the energy remaining in a hydrogen bond after a microscopic displacement of the polymers is transferred to a neighboring bond along the chain, and so on [9]. The original model was further developed by Tolpygo and his collaborators in a series of work; see [10]-[12] and references therein. It is likely that the idea of nearly complete energy recycling in muscles of living organisms can also be applied to explaining a high energy-efficiency of information processing by a human brain, the most energy-efficient computer created so far.

The energy efficiency of electronics, in particular computers, has become a very important problem due to an exponential growth of energy consumption by computational and internet-related systems: supercomputers, data centers, personal computers, etc., which is expected to reach ~ 15% of the total energy consumption in the world in the very near future. Any increase in energy efficiency of electronic systems, or of any area of human activity, would provide tremendous economic and environmental benefits by reducing global warming, achieving sustainable economic development, and protecting the environment. All these topics were of great interest and importance to K.B. Tolpygo, who lectured and published on them profoundly in the later part of his life.

Conventional digital electronics is based on complementary metal-oxide-semiconductor (CMOS) transistor technology where information is encoded by the voltage state of a field effect transistor (FET). The energy dissipation is caused by charging and discharging of the circuit interconnects and gate capacitors of FETs and the gate current leakage in the "OFF" state of transistors. The charging energy is not recycled. Since the invention of the first integrated circuits in the 1960s, semiconductor digital electronics has demonstrated a nearly exponential growth of the integration scale and circuit complexity. The number of transistors per chip has grown by more than eight orders of magnitude, reaching over $1 \cdot 10^9$ (1B) in modern processors and over 20B in field programmable gate arrays (FPGA). At the same time the size of transistors, their gate length, has shrunk from tens of microns down to 14 nm and continues to decrease. Although the gate length has been approaching the physical limits, there is little doubt that semiconductor industry will continue to pack more transistors per chip by using three-dimensional (3-D) integration and other approaches for at least another decade. The progress comes at a higher and higher cost, and the main hurdle is energy dissipation. It has reached ~ 100 W/cm², a factor of 10x higher than the heat density of electric hot plates and induction burners, and a factor of 1000x higher than the solar energy density. Energy dissipation limits the clock frequency of processors to ~ 4 GHz and determines the amount of the so-called "dead silicon," transistors which are not powered at any given time in order to prevent the chip temperature from exceeding the thermal limit.

In the area of high performance computing, the main interest is in advancing supercomputers from the current PFLOPs-scale ($10^{15}$ FLOPS, Floating-Point Operations per Second) to exascale computing, corresponding to $10^{18}$ FLOPs and beyond [13]. A survey of the top 10 supercomputers [14],[15] gives their power consumption in 2015 at ~ 0.1 GW, with the most powerful supercomputer in 2015, Chinese Tianhe-2 (33.9 PFOLPS), consuming about 17.8 MW for operation and another 6.4 MW for cooling. A linear projection from this performance to a 1 EFLOPS (1000 PFLOPS) supercomputer using the same technology gives about 800 MW consumption, the output of an average utility power plant. Current efforts in energy efficiency improvements of supercomputers target a reduction of this figure to about 20 MW by about 2020 [16].

Superconductor electronics (SCE) utilizing Josephson junctions (JJs) as switching devices was historically considered for applications in high-performance computing mainly due to a potential for much higher clock rates (up to a factor of 50x higher) than those offered by the CMOS technology at that time [17]. Recently, superconductor digital electronics has been re-evaluated as having a potential for energy-efficient computing, and energy consumption budgets and other technology requirements have been formulated [16]. A major research program, Cryogenic Computing Complexity (C3), was started in the US in August 2014 in order to develop the technology and demonstrate a prototype of a complete superconducting computer with 10-GHz target clock frequency [18].

Superconductor digital electronics operates at cryogenic temperatures, typically around 4.2 K, and there is currently no technology to enable operation of complex superconducting circuits at significantly higher temperatures. Therefore, the energy required for cryogenic refrigeration must be included in the energy efficiency calculations, which should include the energy dissipation in the circuits as well as all other sources of the heat load such as input/output data and power cables, thermal radiation, etc. Superconductor electronics is also not self-sufficient. A superconducting circuit cannot operate without auxiliary semiconductor electronics such as power supplies, clock generators, output amplifiers, etc. Their energy consumption must also be included in the efficiency calculations.

In order to become competitive with CMOS electronics, superconducting circuits must reach a very large scale of integration (VLSI) that would enable circuit functionalities and complexities required for computing. A result of the author's survey of the Josephson junction count, the simplest measure of circuit complexity, in fully operational superconducting digital circuits, including also JJ-based memory and quantum annealing circuits, fabricated during the last 25 years is shown in Fig. 1.

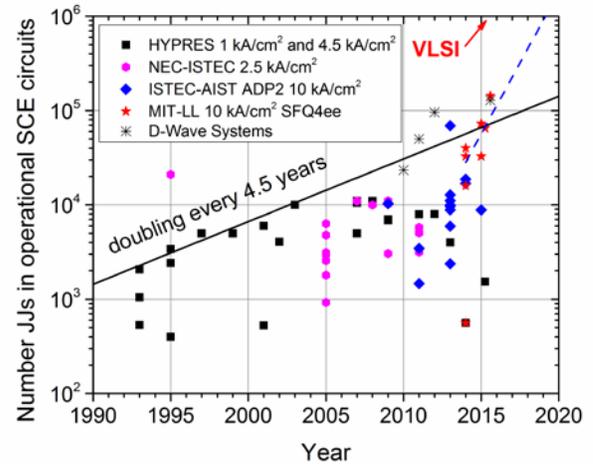

Fig. 1. The total number of Josephson junctions in fully-operational superconducting integrated circuits reported in journal publications and conference proceedings. The circuits were made by the following fabrication processes: HYPRES 1 kA/cm² [19], HYPRES 4.5 kA/cm² [20]-[22], NEC-ISTEC 2.5 kA/cm² standard process [23]-[24], ISTEC-AIST advanced processes ADP [25]-[26] and ADP2 [27], MIT-LL SFQ4ee process [28]-[30], and D-Wave Systems process for quantum annealing processors [31]-[34]. The VLSI boundary corresponds roughly to $10^5$ logic gates or ~$10^6$ JJs. A dashed line shows doubling the number of JJ in circuits every year.

The data represent circuits made in the US and Japan by historically the most successful and currently available fabrication processes for SCE: by HYPRES 1 kA/cm² and 4.5 kA/cm² processes developed at HYPRES, Inc. [35]-[48]; by NEC-ISTEC 2.5-kA/cm² standard process [49]-[64], by ISTEC-AIST 10-kA/cm² advanced process (ADP and ADP2) [26],[53],[61],[65]-[74], by the MIT Lincoln Laboratory 10 kA/cm² process SFQ4ee [75],[76], and by D-Wave Systems, Inc. [31]-[34]. Since any successful integrated circuit is usually a product of a joint work of circuit design and fabrication teams, Fig. 1 characterizes the state of affairs in both the superconducting circuit design and the circuit fabrication areas achieved during the last 25 years. A solid line in Fig. 1 shows an exponential growth with doubling the number of JJs in circuits every 4.5 years. This exponent is a factor of 3 smaller than in the exponential growth demonstrated by CMOS industry during the same period by doubling the number of transistors every 18 months, often referred to as Moore's law. A dashed line in Fig. 1 shows an exponential growth required for achieving goals of the C3 program, doubling the number of JJs per circuit every year.

At present, superconducting digital circuits have about five orders of magnitude lower integration scale than the typical CMOS circuits. E.g., the largest demonstrated Single Flux Quantum (SFQ) circuits have only about $10^5$ Josephson junctions [68],[75],[76] whereas CMOS circuits routinely have over $10^{10}$ transistors. Assuming that all progress in CMOS industry stops right now and superconductor electronics will be capable of sustaining the pace of doubling the number of JJs every year, it will take more than 16.5 years to catch up with the complexity of current CMOS circuits. Several causes of this gigantic disparity have been cited: insufficient funding; lack of profit-driven investments in SCE; immaturity of the fabrication processes and of integrated circuit design tools, etc. If these were the real causes, the corresponding solutions would be trivial: increase funding; develop circuit design tools; use the modern design and fabrication tools; and improve the fabrication process. In the past there have been a few studies predicting the pace of SCE technology development. For instance, in [77] circuits with 1M ($10^6$) JJs were envisioned by 2005, with 10M JJs in 2008, and PFLOPS-scale superconducting computing in 2011. But no one can see the future and obviously none of these predictions turned out to be correct.

In the present work we look a little deeper and, after a brief review of the operation principles and fabrication processes, look at the physical limitations of "classical" SCE and critically analyze whether this technology is energy efficient and scalable to the integration levels required for high-performance computing. We also discuss potential approaches for reducing energy dissipation and increasing the integration scale. Superconducting qubits and numerous problems specific to their implementation in integrated circuits for quantum annealing and gate-based quantum computing will not be considered.

## II. Fabrication and Scalability of SFQ Circuits

Superconducting digital electronics utilizes magnetic-flux (fluxoid) quantization in superconducting loops to define and process bits of information. It is often called SFQ electronics. Nonhysteretic Josephson junctions are used as ultrafast switches. There are three main types of SFQ logic developed to date: Rapid Single Flux Quantum (RSFQ) logic/memory family [78],[79] and its two new 'energy-efficient' versions – ERSFQ [80] and eSFQ [81] – differing from RSFQ only by the biasing schemes; Reciprocal Quantum Logic (RQL) [82]; and Quantum Flux Parametron (QFP) logic [83],[84] and its adiabatic (AQFP) implementation [85],[86].

All superconducting digital circuits present a network of Josephson junctions interconnected by superconducting wires (inductors). DC bias currents are distributed using a network of bias resistors and a common voltage rail in RSFQ circuits or a network of bias inductors and JJs in ERSFQ and eSFQ. Multi-phase AC bias and clock signals in RQL and QFP circuits are distributed using a network of passive transmission lines (PTLs) and coupling transformers. The required switching properties of JJs are achieved using on-chip resistive shunting of hysteretic tunnel Josephson junctions. So, making superconducting integrated circuits reduces to making large networks of JJs, inductors, resistors, and transmission lines.

### A. SFQ electronics fabrication processes

In the semiconductor industry the manufacturing processes are classified by the minimum feature size, which is the gate length of the FETs. Historically, the processes in SCE are classified by the critical current density, $J_c$ of the Josephson junctions, e.g., 10-kA/cm$^2$ process, which is the material property rather than the feature size characteristic. Although the area, $A$ of the junctions is related to $J_c$ as $A = I_c/J_c$, the connection with the process resolution capability is lost since JJs are not the smallest feature in circuits. Also, the ability to route various data and clock signals and interconnect logic cells grows with increasing the number of wiring layers available. So this number and the minimum wiring feature size are also important characteristics. A review of fabrication processes up to 2004 can be found in [87].

At present, there are three advanced fabrication processes for SCE: at the National Institute of Advanced Industrial Science and Technology (AIST) in Japan; at D-Wave Systems, Inc., using Cypress Semiconductor foundry in Bloomington, Minnesota; and at the MIT Lincoln Laboratory. The AIST process was reviewed in detail in [27],[66]. It has two versions: advanced process (ADP) with 10 Nb layers; and ADP2 with 9 Nb layers. Their main limitation is the use of i-line photolithography, which limits the minimum feature size to ~ 0.8 μm, and the use of 3-inch wafers. The D-Wave process has 6 Nb layers, 0.25-μm minimum feature size, and is set on 200-mm wafers. There is no published description of the process, but some information can be found in [31]-[34]. This proprietary process has been used mainly for making quantum annealing processors operating at mK temperatures, which is a very different application than the digital SFQ electronics.

In our recent work [28]-[30], we developed a fabrication process with eight Nb wiring layers and one layer of high-kinetic-inductance material for bias inductors, one layer of Nb/Al-AlO$_x$/Nb Josephson junctions, and full planarization, including the layer of Josephson junctions. So the total number of superconducting layers is nine. This process node was termed SFQ5ee, where "ee" denotes that the process is tuned for making energy efficient circuits for IARPA C3 Program [18]. Here we give a brief review of the MIT-LL fabrication process.

The cross-section of the process is shown in Fig. 2. The target parameters of the layers as they appear in the processing and minimum feature sizes (critical dimensions) are given in Table 1. In comparison with the SFQ4ee process described in [28]-[30], a more advanced SFQ5ee node offers the following enhancements:

**a)** the minimum linewidth and spacing for all metal layers, but M0 and R5, and is reduced to 0.35 μm and 0.5 μm, respectively;

**b)** the minimum size of etched vias and their metal surround is reduced to 0.5 μm and 0.35 μm, respectively;

**c)** the sheet resistance of the resistor layer is increased to 6 Ω/sq by utilizing a nonsuperconducting MoN$_x$ film, offering a choice of either 2 Ω/sq or 6 Ω/sq planar resistors for JJ shunting and biasing;

**d)** an additional thin superconducting layer with high kinetic inductance is added below the first Nb layer M0 in order to enable compact bias inductors;

**e)** an additional resistive layer is added between Nb layers M4 and M5 in order to enable interlayer, sandwich-type resistors with resistance values in the mΩ range for minimizing magnetic flux trapping and releasing unwanted flux from logic cells.

The process consists basically of three main modules: wiring layer module; JJ module; and resistor/kinetic-inductor module. The wiring layer module is shown in Fig. 3.

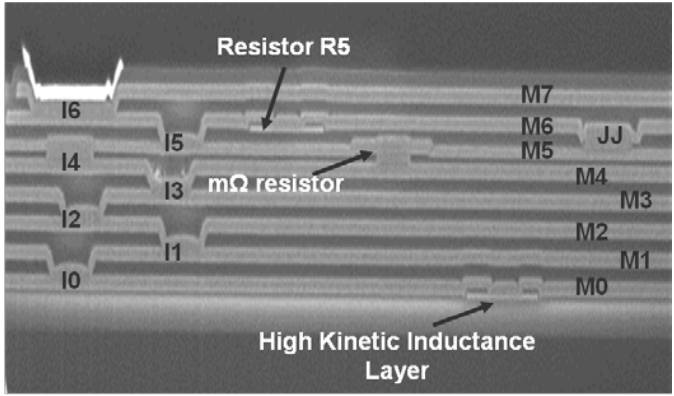

Fig. 2. Focused-ion-beam-made (FIB) cross-section of a wafer fabricated by the SFQ5ee process [30]. The labels of metal layers and vias are the same as in Table I. The additional layers in SFQ5ee process with respect to the previous process node SFQ4ee are: a high-kinetic-inductance layer under M0 and a layer of mΩ-range resistors between M4 and M5 layers.

TABLE I
CRITICAL DIMENSIONS AND LAYER PARAMETERS OF SFQ5EE PROCESS

| Physical layer | Photolithography layer | Material | Thickness (nm) | Critical dimension | | $I_c$ [a] or $R_s$ [b] |
| | | | | Feature (nm) | Space (nm) | |
|---|---|---|---|---|---|---|
| L0 | L0 | $MoN_x$ | 40±10 | 2000 | 500 | 0.5 |
| C0 | C0 | $SiO_2$ | 60±10 | 500 | 500 | |
| M0 | M0 | Nb | 200±15 | 500 | 500 | 20 |
| A0 | I0 | $SiO_2$ | 200±30 | 500 | 500 | 20 |
| M1 | M1 | Nb | 200±15 | 500 | 500 | 20 |
| A1 | I1 | $SiO_2$ | 200±30 | 500 | 500 | 20 |
| M2 | M1 | Nb | 200±15 | 350 | 500 | 20 |
| A2 | I2 | $SiO_2$ | 200±30 | 500 | 500 | 20 |
| M3 | M3 | Nb | 200±15 | 350 | 500 | 20 |
| A3 | I3 | $SiO_2$ | 200±30 | 500 | 500 | 20 |
| M4 | M4 | Nb | 200±15 | 350 | 500 | 20 |
| A4 | I4 | $SiO_2$ | 200±30 | 800 | 800 | 20 |
| M5 | M5 | Nb | 135±15 | 700 | 700 | 20 |
| J5 | J5 | $AlO_x/Nb$ | 170±15 | 700 | 1000 | 100 [c] |
| A5a | I5 | anodic oxide [d] | 40±2 | 700 | 700 | |
| A5b | I5 | $SiO_2$ | 170±15 | 700 | 700 | 20 |
| R5 | R5 | Mo | 40±5 | 500 | 500 | 2±0.3 |
| A5c | C5 | $SiO_2$ | 70±5 | 500 | 500 | 20 |
| M6 | M6 | Nb | 200±15 | 350 | 500 | 20 |
| A6 | I6 | $SiO_2$ | 200±30 | 700 | 700 | 20 |
| M7 | M7 | Nb | 200±15 | 350 | 500 | 20 |
| A7 | I7 | $SiO_2$ | 200±30 | 1000 | 1000 | n/a |
| M8 | M8 | Au/Pt/Ti | 250±30 | 2000 | 2000 | n/a |

[a] $I_c$ is the minimum critical current of metal lines or contact holes (vias) of the critical dimensions (CD) in mA, $R_s$ is the sheet resistance of resistor layer R5 in Ω/sq.

[c] Josephson critical current density of Nb/AlO$_x$-Al/Nb trilayer in μA/μm².

[d] Mixed anodic oxide (AlNb)O$_x$ formed by anodization of Al/Nb bilayer of JJ bottom electrode.

SiO$_2$ was deposited by plasma enhanced chemical vapor deposition (PECVD) at 150 °C.

All wiring layers are processed identically as follows: **a)** Nb layer M$_i$ deposition; **b)** deep-UV photolithography; **c)** high-density plasma etching; **d)** photoresist dry/wet strip. Then metrology steps follow: scanning electron microscopy (SEM) inspection, critical dimension (CD) and thickness measurements. Planarization of the etched metal layer is done by a chemical mechanical planarization (CMP), using the steps shown in Fig. 3: **e)** deposition of a ~ 2.5x times thicker SiO$_2$ over the patterned metal layer; **f)** polishing SiO$_2$ to the required level, using a CMP tool Mirra from Applied Materials, Inc. This is followed by the measurements of the remaining dielectric thickness in 49 points on the wafer, using an elipsometer, and redeposition of SiO$_2$, if needed, to achieve the target ILD thickness in Table 1. Then, the next photography is done on the flat surface of SiO$_2$, Fig. 3(g), in order to etch contact holes through the dielectric to the layer Mi, steps **g)-i)** in Fig. 3. Finally, the next wiring layer M$_{i+1}$ is deposited. This sequence of steps is repeated as many times as the number of wiring layers.

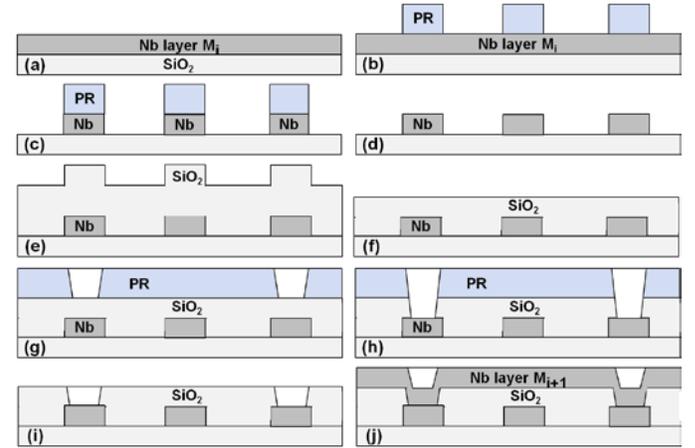

Fig. 3. Processing module of the wiring layers: a) deposition of a wiring layer Mi; b) deep-UV photolithography; c) Nb etching in high-density plasma; d) photoresist dry/wet strip; e) Plasma Enhanced Chemical Vapor Deposition (PECVD) of SiO$_2$ interlayer dielectric for planarization; f) Chemical Mechanical Planarization (CMP) of the interlayer dielectric (ILD) to the required thickness; g) deep-UV photolithography of the interlayer dielectric layer Ii; h) SiO$_2$ etching; i) photoresist dry/wet strip and surface cleaning; j) deposition of the next Nb wiring layer Mi+1. This next Nb layer fills in the etched contact holes in the ILD, thus forming superconducting vias between Nb layers. The sequence of steps is repeated as many times as required by the number of wiring layers.

All metal layers used in the process (Nb, Al, Mo) are deposited on 200-mm Si wafers by dc magnetron sputtering using a multi-chamber cluster tool (Endura from Applied Materials, Inc.) with base pressure of $10^{-8}$ Torr. SiO$_2$ interlayer dielectric (ILD) is deposited at 150 °C, using a Plasma Enhanced Chemical Vapor Deposition (PECVD) system Sequel from Novellus (Lam Research Corporation). Thickness uniformity of the deposited oxide is σ = 2%, where σ is standard deviation (normalized to the mean value). Photolithography is done using a Canon FPA-3000 EX4 stepper with 248 nm exposure wavelength, UV5 photoresist, and AR3 bottom antireflection coating. Etching of all metal and dielectric layers is done in a Centura etch cluster (Applied

Materials, Inc.), using Cl₂/Ar-based chemistry for metals and CHF₃-based chemistry for dielectrics. Etched vias I0, I1, etc. are filled by Nb of the following metal layer.

Josephson junction fabrication was described in detail in [28] and the JJ module is shown in Fig. 4.

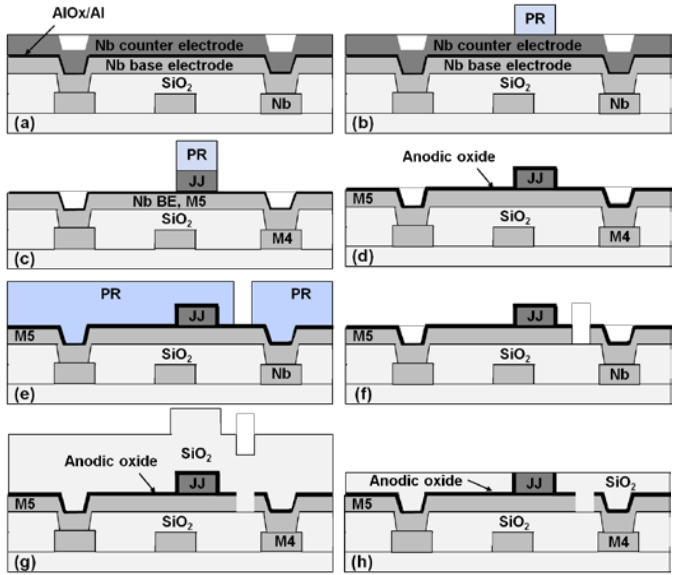

Fig. 4. Josephson junction fabrication module. a) Nb/AlOx-Al/Nb trilayer deposition over the patterned SiO₂ layer. The base electrode of the trilayer fills the etched contact holes, making I4 vias to the bottom Nb layer, M4. b) deep-UV photolithography of the counter electrode to form a junction etch mask; c) junction etching, stopping on AlOx/Al layer; d) anodization; e) photoloithography of the base electrode layer; f) etching of the base electrode, forming wiring layer M5; g) deposition of a thick SiO₂ for planarization; h) CMP to the level of the Josephson junctions to expose their top surface.

The Nb/AlOₓ-Al/Nb trilayer process developed in [88] has been the most successful process for making Josephson tunnel junctions and is used in our work. It consists of a Nb base-electrode deposition (150 nm) followed by in-situ Al deposition (8 nm) and oxidation in pure oxygen at 8 mTorr to achieve the aluminum oxide, AlOₓ, thickness required for 100 µA/µm² Josephson critical current density. A Nb counter-electrode completes the trilayer sandwich. After etching the counter-electrode to define the Josephson junctions, the surface of the tunnel barrier is exposed. To prevent the barrier degradation around the perimeter of the junctions, anodic oxidation is used to form a ~ 50 nm oxide layer on all exposed surfaces, Fig. 4(d). This oxide protects the junctions and allows further processing steps: **e)** photolithography; and **f)** etching of the bottom electrode in order to define Nb wiring layer M5 interconnecting the JJs and connecting them to the bottom layers, Fig. 4. Then, the etched structures are planarized to the level of the tops of JJs as shown by steps **g)** and **h)**.

In order to form resistively-shunted JJs, the following resistor process module is used, Fig. 5. A similar module is used to process the very first layer in the process stack-up, the layer of kinetic inductors, L0, Fig. 1. A resistor layer (Mo or MoNₓ) is deposited on the planarized surface. After the photolithography, resistors are selectively etched in high-density plasma, stopping on Nb and SiO₂. A thin, with ~70 nm thickness, SiO₂ layer is deposited on top to isolate the resistors. Contact holes to the top and bottom electrodes of the junctions and to the resistors are etched. Nb wiring layer M6 is deposited. This M6 layer and layers above it are processed using the wiring module shown in Fig. 3.

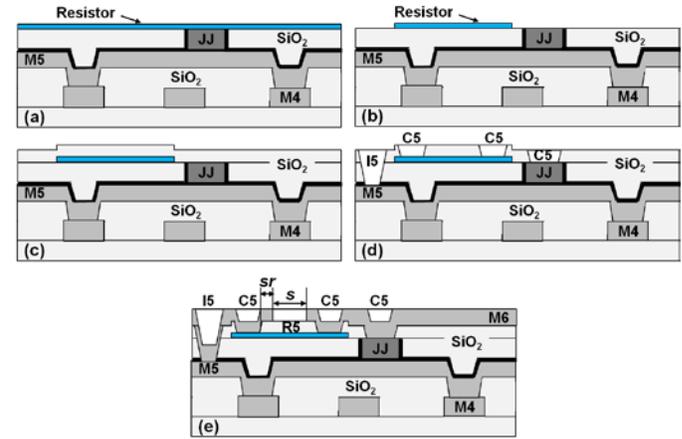

Fig. 5. Resistor/kinetic-inductor process module: a) resistor deposition; b) resistor photolithography, high-density plasma etching and photoresist striping; c) SiO₂ layer deposition, 70 nm thickness; d) photolithography and etching of contact holes to JJ and resistor C5, photolithography and etching of contact holes to the base electrode of JJs, M5; e) Nb wiring layer deposition, M6. Nb layer M6 and layers above it are processed using the wiring module in Fig. 3. The resistor minimum length is determined by the minimum spacing s between superconducting wires and contact holes surround sr shown in (e).

The final process cross section is shown in Fig. 2 and a zoom-in of a cross-section through the junction is shown in Fig. 6. The full 9-superconductor-layer process has more than 400 processing steps.

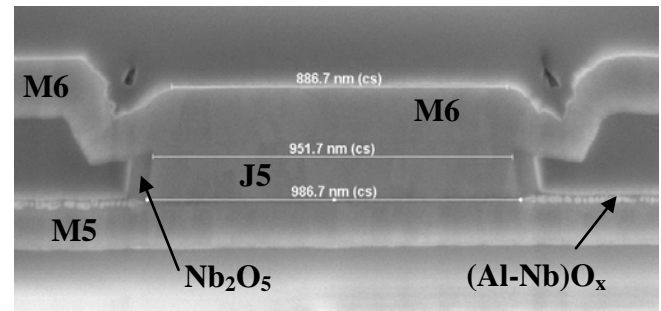

Fig. 6. FIB-made cross-section of a 1-µm Josephson junction. In this particular case, the contact C5 is larger than the junction J5, so the M6 wire overhangs the junction. The opposite situation when C5 is smaller than J5 was shown in Fig. 5. Anodic oxide layer formed on the junction sidewalls and the surface of M5 by anodization is clearly visible.

A perceived simplicity and alleged low cost of the fabrication process were historically cited as one of the main advantages of SFQ electronics [77],[79]. At the time of RSFQ introduction in the US in 1991, the only commercial fabrication process for SCE had only three superconducting niobium layers for interconnecting Josephson junctions, a minimum feature size of 3.5 µm, used Nb/AlOₓ/Nb Josephson junctions, and 3-inch wafers [19]. At that time, these process features corresponded to about Intel's process used to make the semiconductor processors in 1982, and so the process was about 10 years behind, see Table II. The first and very simple RSFQ circuits containing tens of JJs showed great promise and were setting the clock speed records. So, from this starting point, it was tempting to forecast a fantastic growth in future. Since none of the original RSFQ technology proponents was involved with or experienced in integrated circuit technology and manufacturing, it was natural to assign circuit failures to immaturity of the fabrication processes [75] and suggest that their major improvement would be a simple task requiring only very modest investments and second-hand tools from retiring nodes of CMOS manufacturing lines. After 25 years of the SFQ fabrication-technology development, it is clear that these assessments were incorrect. The minimum linewidth of SCE processes has shrunk down to 0.25 μm, the number of superconducting wiring layers has increased from 3 to 9, and the wafer size has increased to 200 mm. So, the features of the currently available SCE processes match and exceed the Intel's process used to manufacture Pentium II processors in 1997, see Table II. However, no SFQ-based computers or digital circuits with complexities comparable to any of the CPUs shown in Table II have emerged.

TABLE II
SEMICONDUCTOR PROCESSORS AND FABRICATION PROCESSES

| Processor | Transistor count | Min linewidth (μm) | No. of metal layers | Year introduced | Chip area (mm²) |
|---|---|---|---|---|---|
| Intel 80186 | 55k | 3.0 | 2 | 1982 | 60 |
| Intel 80286 | 134k | 1.5 | 2 | 1982 | 49 |
| Intel 80386 | 275k | 1.5 | 2 | 1985 | 104 |
| Intel 80486 | 1.2M | 1.0 | 3 | 1989 | 173 |
| Pentium | 3.1M | 0.8 | 3 | 1993 | 294 |
| Pentium Pro | 5.5M | 1.0 | 4 | 1995 | 307 |
| Pentium II | 7.5M | 0.35 | 4 | 1997 | 195 |
| Pentium III | 9.5M | 0.25 | 5 | 1999 | 128 |

## III. PHYSICAL CONSTRAINTS ON VLSI OF SFQ ELECTRONICS

By comparing the features of the SCE fabrication processes described above with CMOS processes given in Table II, one could expect that superconducting circuits with a similar integration scale, similar number of JJs in a few million range, and of similar functionality should be possible to fabricate if the appropriately designed circuits become available. Below we examine this expectation by accounting for the specifics of SFQ circuits.

### A. Josephson Junctions

Firstly, we estimate the maximum possible density of unshunted junctions in SFQ circuits. The total area occupied by a circular junction in M5 (base electrode) or M6 (top wiring) planes is

$$A_J = \pi(r + sr + s/2)^2 = \pi[(I_c/\pi J_c)^{1/2} + sr + s/2]^2, \quad (1)$$

where $J_c$, $r$, $sr$ and $s$ are the Josephson critical current density, the junction radius, base electrode and top wire (M6) surround of the junction, and spacing to the next object, respectively. Then, using $s/2 = sr = 0.25$ μm from Fig. 7 the maximum density of unshunted JJs, $n_J = k/A_J$ as a function of their critical current $I_c$, assuming a 100% area coverage, $k = 1$. In the range of the critical currents typically used, from 100 μA to 300 μA, $n_J$ is from about 15M to 30M JJs per cm². This coverage, of course, is impossible to achieve in a circuit because JJs need to be interconnected to other circuit components. At a more realistic $k$=0.5, $n_J$ is comparable to the density of transistors in the processors in Table II, but three orders of magnitude less than the typical density of modern CMOS transistors. This gap cannot be closed even if the junction technology is pushed to the ultimate values: $J_c$ increased 25 folds to ∼ 2.5 mA/μm² [92]; $sr$ and $s/2$ reduced to 100 nm; and the number of JJ layers increased to two, still giving only $n_J$ ∼ 5·10⁸ JJ/cm².

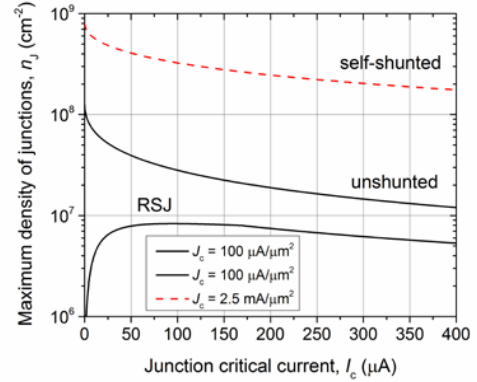

Fig. 7. The maximum density of Josephson junctions in SFQ circuits as a function of the average $I_c$ of the SFQ cells. Resistively-shunted and unshunted JJs in the current technology node SFQ5ee ($J_c = 0.1$ mA/ μm², $sr = s/2 = 0.25$ μm) are shown by the solid lines; self-shunted JJs in a hypothetical technology node with $J_c = 2.5$ mA/μm² and $sr=s/2= 0.1$ μm are shown by the dashed line. A 100% area coverage is assumed, $k = 1$.

SFQ circuits utilize nonhysteretic junctions. In the existing technology this is achieved by resistive shunting of the tunnel junctions as shown in Fig. 5. The top view of a resistively shunted junction (RSJ) is shown in Fig. 8. The total area of the RSJ includes the junction area $A_J$, resistor area, and the area of vias and wires connecting the JJ and the resistor. This area includes also the overlap between wires M5 and M6 and the contact holes and the junction by amount $sr$, a process parameter given in the design rules document. Below we estimate the total RSJ area in order to estimate the maximum circuit density.

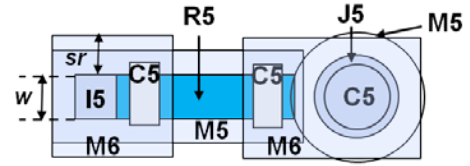

Fig. 8. Top view of a resistively-shunted JJ showing JJ counter electrode J5, JJ bottom electrode M5, resistor R5, contact holes C5 and I5, and wiring layer M6 providing connections between the JJ and the resistor. All metal wires M5 and M6 must overlap (surround) the JJ and all contact holes by some amount, $sr$ according to the process design rules.

SFQ circuits utilize RSJs with critical damping, i.e., with McCumber-Stewart [89],[90] parameter $\beta_c = 2\pi I_c R_n^2 C/\Phi_0 \approx 1$, where $\Phi_0 \equiv h/2e$ is the flux quantum, $C$ is the junction capacitance, $R_n$ is the damping resistance assumed to be a parallel combination of the junction internal resistance $R$ and the shunt resistance $R_s$, see inset in Fig. 9. The inductance $L_s$ associated with the superconducting connections to the shunt and of the shunt itself makes damping frequency-dependent, which is usually neglected in SFQ circuit design. The internal resistance is usually approximated by a piecewise function:

$$R = R_{sg} = \gamma R_N \text{ for voltages } V < V_g \quad (2a)$$

$$R = R_N \text{ at } V \geq V_g, \quad (2b)$$

where $V_g = 2\Delta/e$ is the gap voltage in the symmetrical tunnel junction, $\Delta$ is the energy gap in the electrodes, $R_N$ is the normal state tunnel resistance, $R_{sg}$ is the subgap resistance, and $\gamma$ is the temperature- and JJ-quality-dependent coefficient, $\gamma > 1$. Then, neglecting $L_s$, the damping resistance at low voltages $V < V_g$ is $R_n = \gamma R_N R_s/(\gamma R_N + R_s)$.

The typical current-voltage characteristics of JJs in the MIT Lincoln Laboratory process using Nb/Al-AlOₓ/Nb tunnel junctions with Josephson critical current density of 10 kA/cm² (100 μA/μm²) are shown in Fig. 9 for the unshunted junction (curve 4) and junctions of the same size with three different values of the shunt resistor corresponding the characteristic voltage $V_c \equiv I_c R_n$ of 0.30 mV (curve 1), 0.69 mV (curve 2), and 0.96 mV (curve 3), and to $\beta_c = 0.2$, 1, and 2, respectively.

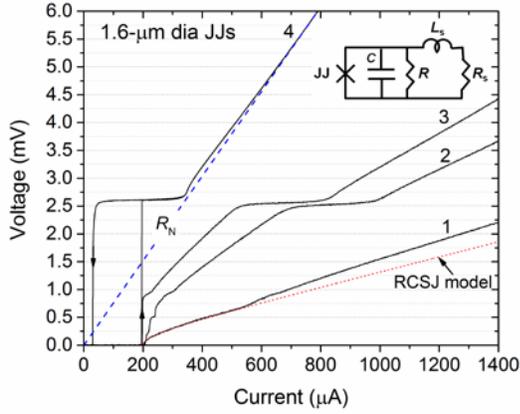

Fig. 9. Current-voltage characteristics of resistively-shunted Josephson junctions in the MIT-LL fabrication process SFQ5ee with $J_c = 100$ μA/μm². Data for 1.6-μm-diameter junctions with three different shunt resistors $R_s$ are shown: (1) $R_s = 1.6$ Ω; (2) $R_s = 3.73$ Ω; (3) $R_s = 5.4$ Ω, corresponding to $V_c \equiv I_c R_n$ values of 0.30 mV, 0.69 mV, and 0.96 mV, respectively. These values in turn correspond to $\beta_c = 0.2$, 1, and 2, respectively. The top $I$-$V$ curve (4) is for the unshunted JJ. Inset shows the circuit diagram. The shunt inductance associated with the current path from the JJ along the resistor to the I5 via and back to the JJ along the M5 electrode, $L_s \sim 1$ pH, is usually neglected. However, it causes an internal $L_s$-$C$ resonance with the junction capacitance, corresponding to a step at ~ 0.5 mV in curve (2).

At the specific capacitance value of 70 fF/μm² given in the SFQ5ee process design rules, the characteristic voltage of the junctions at $\beta_c = 1$ is $V_c \equiv I_c R_n \approx 686$ μV. At $J_c = 100$ μA/μm² used in the process, $R_{sg} >> R_s$, $R_N$, $\gamma \approx 10$ in (2), and its contribution can be neglected in the estimates of the shunt resistor here. Then, the shunting resistor value is simply $R_s = V_c/(J_c A) = 686/I_c$, where $R_s$ is in ohms and $I_c$ in μA. The area of a resistor $R_s = R_{sq} l/w$ with the minimum linewidth $w$ and length $l$ depends on the sheet resistance of the material used, $R_{sq}$ and other process parameters, see [30],

$$A_R = w^2 R_s/R_{sq} \text{ if } R_s/R_{sq} \geq (s + 2 \cdot sr)/w \quad (3a)$$

$$A_R = (s + 2 \cdot sr)^2 R_{sq}/R_s \text{ if } R_s/R_{sq} < (s + 2 \cdot sr)/w, \quad (3b)$$

because the resistor length cannot be made shorter than $l_{min} = s + 2 \cdot sr$, about 1 μm in the current process node. At $R_{sq} = 2$ Ω/sq, the boundary corresponds to $R_s = 4$ Ω. Therefore, all JJs with $I_c \geq 172$ μA have shunts in the regime (3b). We need to add the area of two C5 and one I5 vias with surround, which is approximately $3(w + 2 \cdot sr)^2$ in the regime (3a) and $(w + 2 \cdot sr)^2 + 2(s + 2 \cdot sr)(w + 2 \cdot sr)R_{sq}/R_s$ in the regime (3b), and account for the spacing to the next feature, where $w$ is the minimum size of features, see Table I.

Then the total area of an RSJ becomes:

$$A_{RSJ} = \pi[(I_c/\pi J_c)^{1/2} + sr + s/2]^2 + w(w + s)V_c/(I_c R_{sq})$$
$$+ 3(w + 2 \cdot sr + s)^2, \quad (4a)$$

for $I_c < [w/(s + 2sr)]V_c/R_{sq} \approx 172$ μA, and

$$A_{RSJ} = \pi[(I_c/\pi J_c)^{1/2} + sr + s/2]^2 + (s + 2 \cdot sr)[s + 2 \cdot sr + (s + 2 \cdot sr)I_c R_{sq}/V_c]$$
$$+ (w + 2 \cdot sr + s)^2 + 2(w + 2 \cdot sr + s)[s + (s + 2 \cdot sr)I_c R_{sq}/V_c] \quad (4b)$$

for $I_c \geq 172$ μA.

The maximum possible density of RSJs $n_{RSJ} = 1/A_{RSJ}$ following from (4) is plotted in Fig. 7, bottom curve. It is significantly lower than the maximum density of unshunted junctions. The maximum density is about 8.3M RSJs per cm² and nearly independent of the critical current of JJs in the range from ~ 70 μA to ~ 175 μA. The RSJ area in this range is $A_{RSJ} \sim 12$ μm². The maximum density of RSJs in SFQ circuits can be estimated as $k/A_{RSJ}$ by using in (4) the most frequently encountered, or the average, critical current $<I_c>$ and the area filling factor $k \sim 0.5$. Inspection of all RSFQ, ERSFQ, etc. cells in [77]-[81],[91] shows that $<I_c> \approx 175$ μA. It is a result of selecting $I_c \approx 100$ μA as the minimum value used in the cells, based on the maximum acceptable bit error rate. Since the junction must be connected to inductors, the RSJ area coverage of 25% to 50% is more realistic, reducing the maximum circuit density to about 2M to 4M RSJs per cm².

### B. Statistical Variations of Josephson Junctions

As was shown above, the maximum density of RSJs is nearly independent of the choice of $<I_c>$ and sufficient to place a few million of JJs on a 1-cm² chip. However, it is important to check if they all can be yielded with critical currents within the required margins. From the circuit standpoint, statistical variations of JJ critical currents, as well as thermal and quantum noise, induce storage, decision, and timing errors and determine the bit error rate in SFQ circuits [93]. From the fabrication process standpoint this is related to the so-called parametric yield - the fabrication yield of devices with parameters lying within a given range ±$M$ with respect to the targeted mean value. The SFQ cells are designed to tolerate some relatively large deviations. For instance, all critical currents of the junctions can be changed by about ±30% if changed uniformly, or any single junction can deviated from the target by the ±30% if all other JJs are on target, etc. However, random deviations of all junctions and all inductors in the cells cause significant margin shrinkage. The typical bias margins reported for the circuits shown in Fig. 1 are often less than ±10% especially at high clock frequencies.

The statistics of JJ critical currents fabricated by our planarized process was studied in [28] and can be described as approximately Gaussian with the standard deviation depending strongly on the junction size. This dependence comes from the fluctuations of the junction area caused by photolithography and tunnel barrier transparency fluctuations, both increasing with decreasing the junction diameter. Statistical data in [28] can be converted into the dependence of the standard deviation of $I_c$ (normalized to the mean value) on the critical current (in µA) as

$$\sigma_I = 0.115 I_c^{1/2}/(I_c - I_0), \qquad (5)$$

where the dimensional prefactor is in $\mu A^{1/2}$ and $I_0$ is the critical current cut-off – the process resolution characteristic corresponding to the critical current of the smallest resolvable junction. In the SFQ5ee process using 248-nm photolithography, this minimum size is about 250 nm and $I_0 \approx 5$ µA.

The probability that a JJ critical current is within the circuit margins ±$M$ is given by the error function $p = erf(M/\sqrt{2}\sigma_I)$. For an $N$-junction circuit, the probability (yield $Y$) that all JJs are within ±$M$ range is $Y = p^N$. Then, the fabrication-yield-limited maximum number of JJs in a circuit is

$$N_{max} = \ln(Y_{min})/\ln[erf(M/\sqrt{2}\sigma_I)], \qquad (6)$$

where $Y_{min}$ is the minimum acceptable circuit yield. This dependence at $Y_{min} = 50\%$ and 90% is plotted in Fig. 10 for three different values of the circuit margins $M$: 15%, 10% and 5%. Also shown is the maximum number of RSJ which can be placed on a chip with area $A_{ch}$

$$N_{RSJ} = kA_{ch}/A_{RSJ} \qquad (7)$$

at fill factor $k = 0.5$ and $A_{ch} = 1$ cm² and 2 cm².

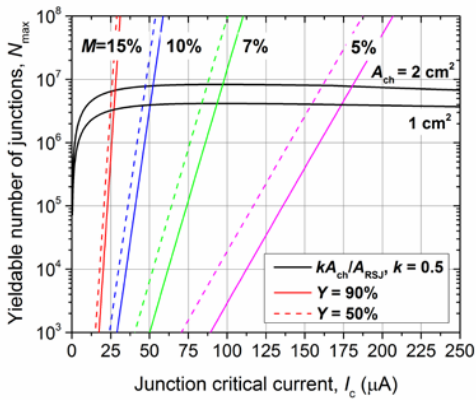

Fig. 10. The maximum number of Josephson junctions which can be yielded with a required probability (circuit yield) $Y$ by the current fabrication process SFQ5ee and with all junctions having the critical current within the circuit margins ±$M$, as given by (6) and (5). Dashed lines and solid lines correspond to $Y = 0.5$ and $Y = 0.9$, respectively. Also shown is the maximum number of resistively-shunted JJs $kA_{ch}/A_{RSJ}$ which can be placed on a chip with 1 cm² and 2 cm² area at 50% area coverage, $k = 0.5$, and $J_c = 100$ µA/µm².

At critical currents smaller than the point of intersection of (6) and (7) the maximum number of JJs in the operational circuits is determined by the $I_c$ spreads, acceptable circuit yield, and the circuit margins. At higher critical currents, this number is limited only by the junction area and can be increased by increasing the size of the chip. It can be seen that at $<I_c> \approx$ 175 µA and above, as was chosen in RSFQ originally, the size of the circuit (JJ count) is not limited by the parametric yield of the current process even for poorly designed circuits with ±5% margins, and the circuit parametric yield above 50% can be reached even on large-area chips ~ 2 cm² with about 10M RSJs. However, the current desire for energy efficiency requires reducing $<I_c>$ well below this number, see Sec. IV. Below $<I_c> = 50$ µA, a value popular in many RQL and AQFP designs and used in [16] for calculating SCE electronics power budgets, the circuit complexity will be limited by the fabrication process unless circuits with very wide margin can be designed.

It is author's experience, however, that the practical yield of complex circuits is much lower than the parametric yield following from the assumed normality of the parameter distribution used in our estimates here. In practice, the circuit yield is determined by defects and outlier devices, i.e., devices in the far tails of the distribution. The probability of outliers increases with decreasing the junction size. The distribution is often skewed and the tails are usually nongaussian. Often they can be described by the Weilbull statistics with a simple exponential decay. A more detailed description of this subject is beyond the scope of this work and will be presented elsewhere. However, the message of this is that decreasing $<I_c>$ below ~ 75 µA is expected to compromise the circuit yield in the current technology node.

## C. Limitations on Scaling Caused By Bias Currents

RSFQ-based circuits, including ERSFQ and eSFQ, use a parallel dc biasing of JJs from a common voltage rail. The typical bias current is $0.7I_c$. The total bias current $I_{tot}$ grows proportionally to the number of JJs and can be estimated as

$$I_{tot} = 0.7 N_{RSJ} <I_c>, \qquad (8)$$

where $N_{RSJ}$ is the number of JJs in the circuit, giving about 122 A per 1M JJs. It is clear that such large currents cannot be supplied to and handled by thin-film superconducting layers at 4 K. Large bias currents also create large stray magnetic fields and cause magnetic flux trapping and circuit margin degradation. The maximum current which has successfully been delivered to the largest operational RSFQ circuit is ~ 3 A [69], with a substantial margin degradation of some of its sub-circuits. This total-bias-current limitation caused the saturation in the number of JJs in RSFQ circuits at ~ 12 thousand (12k) JJs, which can be seen in Fig. 1 during a 10-year period from ~ 2004 to 2014. All circuits with JJ count over 20k shown in Fig. 1 used various ac serial-biasing schemes.

RSFQ-based circuits or any circuits with parallel biasing are not scalable beyond about 20k to 30k JJs. To mitigate this problem, serial biasing of RSFQ circuits was proposed a long time ago [94] and demonstrated in relatively simple circuits in [95]-[97]. Serial basing, also known as current recycling, requires breaking a circuit into $m$ isolated islands and recycling the return current from the ground planes of one island to bias in series the next island, thus reducing the total current by a factor of $m$. The current drawn by each island must be equal, and the input and output currents must not add currents to the serially biased circuits. This is achieved by transferring microwave clock and SFQ data between the islands through transformers using driver-receiver pairs. For an $m$-bit processor, a natural recycling scheme would be between the $m$ individual bits. Since the number of JJs per island should be kept at ~ 10k level in order to keep the total current below 2 A, the number of islands in a 2M-junction circuit becomes >100, requiring lots of inter-island interface circuitry with its own biasing. This complicates design and decreases the circuit density. Presently, there are no readily available and proven solutions for making VLSI SFQ circuits with current recycling, and many technical problems remain to be solved.

RQL and QFP circuits use multiphase ac currents for clock and bias, so the total current only weakly increases with $N_{RSJ}$. The technical difficulties with ac biasing lie in the proper distribution of these high-frequency currents along PTLs to each junction and the negative effect on the circuit density caused by the PTLs and coupling transformers.

### D. Heating of Resistors

Many SFQ circuits use resistors and resistive dividers for distributing dc bias currents and matching RF impedance. These resistors are in direct contact with superconducting wires on M6 layer through C5 vias. Heat dissipated in the resistors increases the local temperature and decreases the critical current of wires and JJs, and in extreme cases can turn them into the normal state. This should be mitigated by a proper sizing of resistors and wires, and proper spacing of high-current-carrying resistors from other circuit elements sensitive to temperature increase.

The amount of heat dissipated in a thin-film resistor with sheet resistance $R_s$, width $w$, and length $l$ per unit time is

$$Q_{diss} = I^2 R_s l/w, \qquad (9)$$

where $I$ is the current. In the steady state, this amount of heat power is balanced by the heat conduction through the circuit layers into the substrate and into liquid helium (or a cryocooler) cooling the chip surfaces. The amount of heat that can be removed due to conduction can be estimated as

$$Q_{cond} = A_R \Delta T / R_{th}, \qquad (10)$$

where $A_R = 2lw$ is the resistor surface area, $R_{th}$ is the effective thermal resistance, and $\Delta T$ is the resistor temperature increase. Equating (9) and (10) we get

$$\Delta T = I^2 R_{sq} R_{th} / (2w^2), \qquad (11)$$

i.e., the resistor temperature increases inversely with the resistor width squared. There is a maximum temperature increase $\Delta T_{max}$ above which Nb wires in contact with the resistor will transition into the normal state at a given current and given bath temperature. This sets the maximum value of the current in the resistor as

$$I_{max} = w[2\Delta T_{max}/(R_{sq}R_{th})]^{1/2}. \qquad (12)$$

The maximum current one can supply through the resistor without turning the contacting Nb wires into the normal state is inversely proportional to the square root of the sheet resistance. So the desire to increase $R_{sq}$ in order to minimize the resistor area, see (3a), is at odds with the resistor heat handling capabilities. Since heating of SFQ circuits is highly undesirable, (11) and (12) put strong restrictions on the minimum width of resistors that can be used or the currents they can handle, strongly impacting the integration scale.

The thermal resistance at both interfaces of the resistor is not exactly known. For a metal surface in contact with liquid helium at 4.2 K, the typical thermal resistance $R_{th}$ is about 1 K $\mu m^2/\mu W$. Using this value of $R_{th}$, $R_s = 2 \, \Omega/sq$, and $\Delta T_{max} = T_{cNb}$ - 4.2 K = 5 K we estimated the maximum current per unit width of the resistor as $I_{max}/w = 2.2$ mA/$\mu$m. Simple measurements of $I$-$V$ characteristics of Nb wires in contact with Mo resistors of different width done in this work give $I_{max}/w = 1.3$ mA/$\mu$m. This value translates into the effective thermal resistance of $R_{th} \approx 3$ K $\mu m^2/\mu W$ ($3 \cdot 10^{-6}$ K m$^2$/W) for the circuit resistors buried deep into the multilayered structure with multiple interfaces between Nb and SiO$_2$ layers (Fig. 1).

Resistor heating makes it almost impossible to achieve very large integration scale in circuits with parallel biasing. For instance, distributing 244 A bias current in a circuit with 2M JJs, considered in Sec. IIC, would require a total width of bias resistors of ~ 20 cm.

### E. Circuit Inductors

Yet another constraint comes from the circuit inductors, since every JJ in an SFQ circuit is connected to an inductor. By inspecting the published RSFQ and RQL cells and circuits, we note that the average superconducting loop with JJs in the designs has a dimensionless parameter $\beta_L = 2\pi I_c L/\Phi_0 \sim 2$, where $L$ is the inductance. We denote this average value as $\langle\beta_L\rangle$. Each inductor occupies area $A_L = (L/\ell)(w+s)$, where $\ell$, $w$ and $s$ are the inductance per unit length, inductor linewidth, and spacing between the inductors, respectively. The maximum inductor density grows linearly with the average critical current of SFQ cells as

$$n_L = 2\pi \ell \langle I_c \rangle m_L k / [\Phi_0 (w+s) \langle\beta_L\rangle], \qquad (13)$$

where $m_L$ is the number of physical layers of inductors, $k$ is the filling factor. In order to minimize cross-talk between inductors, stripline configuration is used predominantly. This requires three superconducting layers: one signal layer and two ground planes. So, a process with eights superconducting layers may have up to three completely independent layers of inductors. The smallest inductor linewidth in our process is 0.35 $\mu$m and spacing is 0.5 $\mu$m, so $w+s=0.85$ $\mu$m. At this linewidth, the typical stripline inductance per unit length is $\ell \approx 0.6$ pH/$\mu$m [29]. Since, on average, each RSJ requires an inductor, the maximum circuit density can be estimated by equating $n_{RSJ}$ and $n_L$.

The plot of (13) is shown in Fig. 11, along with $n_{RSJ}$ and $n_J$ as a function of the average critical current $\langle I_c \rangle$, for a few values of $\langle\beta_L\rangle$ and $m_L$, and $k = 0.5$. (It should be noted that some fraction of the area is occupied by vias providing connections between the inductors and the junctions, which may reduce the actual value of $k$ below 0.5.) We can see that circuits with two layers of inductors can reach the maximum of about 3M components (both RSJs and inductors) per 1-cm$^2$ chip if $<I_c>$ is larger than about 35 µA. Circuits with just one layer of inductors may still reach the same complexity if $<I_c>$ is larger than ~ 75 µA. Inductance approach has been taken in designing AC-biased shift registers [75], and circuits with densities over 0.6M RSJs per cm$^2$ have been demonstrated. Recently these shift registers have been redesigned and fabricated at MIT-LL by the SFQ5ee process. Operational circuits with 1.3M RSJs per cm$^2$ density and over 65000 RSJs have been demonstrated [99]; see Fig. 12. Testing of shift registers with 144000 RSJs is the work in progress.

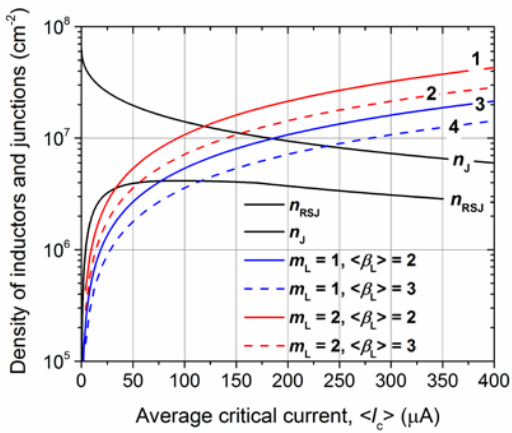

Fig. 11. The maximum density of circuit inductors in the SFQ5ee process as a function of the average critical current in SFQ cells $<I_c>$, assuming the average $<\beta_L> = 2$ and $m_L = 2$ (curve 1), where $m_L$ is the number of independent layers of inductors. Also shown are: $<\beta_L> = 3$, $m_L = 2$ (curve 2); $<\beta_L> = 2$, $m_L = 1$ (curve 3); and $<\beta_L> = 3$, $m_L = 1$ (curve 4); the density of shunted and unshunted junctions $n_{RSJ}$ and $n_J$. Inductance per unit length of 0.6 pH/µm [29] was used, and the area fill factor $k = 0.5$ was assumed.

At $<I_c>$ lower than the intersection point of the $n_L(I_c)$ and $n_{RSJ}(I_c)$ dependences, the circuit complexity is limited by the number of inductors and, at higher $<I_c>$, by the number of junctions. Circuits with $<I_c>$ larger than about 125 µA are not limited by inductors at all, only by the area occupied by RSJs.

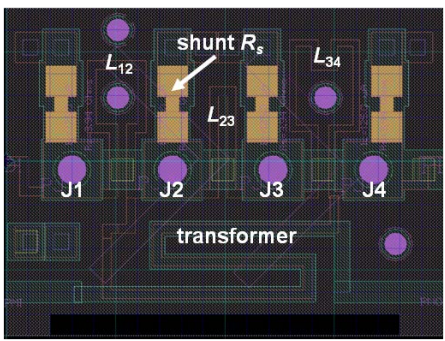

Fig. 12. An example of the unit cell (bit) of the AC-biased shift registers [99]. The cell dimensions are 20 µm x 15 µm, the RSJ density is $1.3 \cdot 10^6$ RSJs per cm$^2$. A 16363-bit fully operational circuit has 65536 RSJs. The minimum linewidth, $m_L$, and $I_c$ used were 0.4 µm, 1, and 125 µA, respectively.

It is also clear from Fig. 11 that increasing $m_L$ beyond 2 by increasing the number of superconducting layers in the process will not increase the circuit density, because it is limited by the density of RSJs. However, more layers may add more flexibility in routing the clock, bias, and data paths.

It is interesting to note that, if the resistive shunts could be eliminated by using self-shunted JJs with the same $J_c$ and the similarly tight parameter spreads as the shunted tunnel junctions, the circuit complexities can reach about 10M (JJ/inductors) per 1-cm$^2$ chip by choosing $<I_c>$ in the range from ~120 µA to 190 µA.

## IV. ENERGY DISSIPATION AND EFFICIENCY OF SCE

### A. RSFQ-based and RQL circuits

RSFQ logic was proposed more than 30 years ago as a replacement of Josephson latching logic used by IBM in its early attempt to build a Josephson-junction-based computer in the 1970s – early 1980s. In RSFQ logic/memory the information is encoded by the presence (logic 1) or absence (logic 0) of a single flux quantum $\Phi_0 \equiv h/2e$ in a logic cell and is transferred between the cells in the form of picosecond-wide SFQ voltage pulses. An SFQ pulse is a voltage pulse generated across a Josephson junction when the phase difference across the junction flips by $2\pi$ as a result of some external perturbation, e.g., a current pulse. The second Josephson relation [100] $d\varphi/dt = (2e/\hbar)V$ guarantees that these SFQ pulses have a quantized area equal to a single flux quantum $\int V(t)dt = \Phi_0 \approx 2.07$ mV·ps or 2.07 pH·mA. The process of SFQ pulse generation can be also viewed as a passage of a flux quantum through the junction. Accordingly, if a junction is embedded into a superconducting loop, such a passage also changes (reduces or increases) the flux through the loop by $\Phi_0$. A complete description of SFQ logic was given in [79] and the cell library can be found in [91]. Each cell has separate data inputs and outputs, and clock lines. SFQ pulses encoding data and clock are distributed on two different networks of JTLs and PTLs. In the time domain, logic "1" is encoded by the arrival of a data SFQ pulse (on the data input) between two clock pulses (on the clock input), whereas logic "0" corresponds to no data pulse between the clock pulses.

To provide reliable switching by an SFQ pulse and set the direction of SFQ pulse/flux propagation, almost all JJs in the circuits are current-biased at a value $I_b \approx 0.7I_c$, using a network of either bias resistors and a common voltage rail (as in RSFQ) or a network of inductors and current-limiting junctions (as in ERSFQ, eSFQ), or their combinations.

The average dynamic-power dissipation in an SFQ circuit utilizing Josephson junction switching (RSFQ, ERSFQ, eRSFQ, RQL) can be estimated as

$$P_{cold} = \alpha N f_{cl} <E_{sw}>, \qquad (14)$$

where $N$ is the number of Josephson junctions in the circuit, $f_{cl}$ is the clock frequency, $\alpha$ is the activity coefficient, the fraction of JJs switching during the clock period, $<E_{sw}>$ is the average energy loss per switching.

It is well known that a resistively capacitively shunted junction (RCSJ) has a potential energy described by a tilted washboard potential:

$$E(\varphi) = E_J(1-\cos\varphi - i\varphi), \qquad (15)$$

where $E_J = I_c\Phi_0/(2\pi)$ is the Josephson energy and $i = I_b/I_c$ is the normalized bias current. When switched by an SFQ pulse, the junction goes from one potential minimum of (15) to another separated by a $2\pi$ phase difference. The energy difference between the two minima is $\Delta E = 2\pi i E_J$. Since the junction is critically damped or overdamped, all of this energy is dissipated in the resistor, and there is no energy left for recycling and increasing the efficiency of the information processing. Hence, the average energy loss per switch is

$$<E_{sw}> = <I_b>\Phi_0, \qquad (16)$$

where $<I_b>$ is the average bias current. This energy should not be confused terminologically with the switching energy – the energy required to flip the junction's phase by $2\pi$. The latter equals to the height of the potential barrier between the two adjacent minima of (15) and can be made arbitrarily small by increasing the bias current.

Using the typical value $<I_b> = 0.7<I_c>$, following from the circuit speed optimization, and assuming a random mix of "ones" and "zeros" ($\alpha \approx 0.5$), the average power loss in an RSFQ-type circuit is

$$P_{cold} = 0.35 N f_{cl} <I_c> \Phi_0. \qquad (17)$$

The RQL circuits use four-phase ac currents for JJ biasing and circuit clock [82]. A reciprocal pair of SFQ pulses (positive and negative) during the clock period encodes "1" and no pulses encode "0". This encoding increases the energy loss by a factor of two, and for a random mix of "ones" and "zeroes" results in $\alpha = 1$ in (14). Also, according to [82], the switching of JJs by an ac bias current occurs at a current smaller than $I_c$. As a result, the energy dissipation in RQL circuits with random data can be approximated by

$$P_{cold} = (1/3) N f_{cl} <I_c> \Phi_0, \qquad (18)$$

where $<I_c>$ is the weighted average of the critical current of the junctions [82]. The difference between (18) and the RSFQ-based case (17) is completely negligible.

Inspection of all RSFQ cells in [77]-[79], [91] shows that the average critical current $<I_c> = 0.175$ mA. This gives the average energy loss $<E_{sw}> = 2.5 \cdot 10^{-19}$ J or 0.25 aJ per JJ switching in RSFQ, ERSFQ, and eSFQ circuits. This is a factor of $6 \cdot 10^3$ times larger energy loss than the Landauer's minimum energy-per-bit requirement for irreversible computing, $k_B T \ln 2$ [101] at $T = 4.2$ K; $k_B$ is the Boltzmann constant. It is important to stress that the minimum critical current used in the SFQ cells is typically a factor of two lower than the average as a wide range of JJs values is used in the cells. However, the minimum critical current sets the maximum acceptable bit error rate [77],[79] and typically is about 0.1 mA. Changing the minimum $I_c$ value by a factor of $m$ would automatically change the $<I_c>$ and the $<E_{sw}>$ by the same factor.

Another important quantity is the energy loss per a single-bit operation (SBOP). It is different from $<E_{sw}>$ because any logic gate (Boolean) operation requires switching of multiple JJs, so

$$<E_{SBOP}> = N_{SBOP} <E_{sw}> , \qquad (19)$$

where $N_{SBOP}$ is the average number of JJ switches required. This number in RSFQ and RQL cells is about 10. For instance, OR gate has 12 JJs, XOR gate has 9 JJs, AND gate 11 JJs, etc. [77],[79],[91]. Recall that three SFQ pulses (3 switches) are required just to encode "1" and two switches to encode "0". So, $<E_{SBOP}> \approx 10<E_{sw}> \approx 2.5$ aJ.

In order to compare the energy efficiency of the cryogenic electronics with room-temperature electronics we need to account for the energy loss associated with cryocooling. Removing $P_{cold}$ from the chip at 4.2 K requires a cryocooler (a heat machine) consuming a much larger power, $P_{hot}$ at 300 K, given by

$$P_{hot} = P_{cold} / \varepsilon = P_{cold} (T_{hot} - T_{cold}) / (\eta T_{cold}), \qquad (20)$$

where $\varepsilon = \eta \varepsilon_{id}$ is the energy efficiency of the cryocooler and $\varepsilon_{id} = T_{cold}/(T_{hot} - T_{cold}) \approx T_{cold}/T_{hot}$ is the efficiency of an ideal thermal machine (Carnot efficiency) and $\eta < 1$ is a nonideality factor. Depending on the cryocooler size and type, this factor changes from $\eta \sim 0.02$ for the small-scale ($\sim 1.5$ W at 4.2 K) pulse-tube coolers to $\sim 0.20$ for the largest-scale Linde helium liquefiers with $\sim 400$-kW wall power requirements, see e.g., Table 1 in [16]. The inverse efficiencies of these two types of cryocoolers, $1/\varepsilon_{sm} = 3520$ and $1/\varepsilon_l = 352$, are used hereafter for all power consumption estimates as the upper and the lower boundaries.

In order to determine which technology is more energy efficient, CMOS or SFQ, we need to compare the power loss in two circuits performing a similar function or a similar amount of information processing. The performance and power loss in the CMOS-based processors can easily be measured, are well known and can be found in [102]. The power dissipation is typically below 140 W. For instance, Intel's quad-core Core i7-4790 Haswell CPU using 22-nm technology node has 88-W power dissipation at $f_{cl} = 4$ GHz, performance of about 200 GFLOPS, and $N = 1.4 \cdot 10^9$ transistors [103]. Unfortunately, superconducting processors of comparable complexity do not exist, and the achieved integration scale differs by 5 orders of magnitude. So, to make a comparison, we need to make some assumptions. Below, we provide a few estimates of the upper and lower bounds on the power consumption in VLSI SFQ circuits.

Firstly, we note that the number of JJs in RSFQ and RQL logic gates and memory cells is comparable or even larger than in CMOS logic gates and memory. Secondly, superconducting processor architectures that are being developed use algorithms developed for CMOS computers and emulate their architectures, using adders, multipliers, etc., see, e.g., [47], [63]-[67], [71]-[74], [104]. It is reasonable to assume then that a superconducting processor with $N$ junctions operating at $f_{cl}$ will be processing about the same amount of information (perform the same logic and memory functions) as a CMOS-based processor with $N$ transistors clocking at the same frequency. Then, using $f_{cl} = 4$ GHz, $N = 1.4 \cdot 10^9$ JJs, and $<E_{sw}> = 2.5 \cdot 10^{-19}$ J estimated above for $<I_c> = 0.175$ mA, we get from (17) and (18) $P_{cold} = 0.71$ W, see Table III. This is a low power compared with about 100 W power consumed by the CMOS chip operating at room temperatures.

However, depending on the cryocooler used, the total power consumption by our hypothetical SFQ circuit is from $P_{hot} \approx 250$ W to 2.5 kW, a factor of $\sim 3x$ to 30x larger than in the CMOS processor of the same complexity. This, perhaps, is a rather surprising result for many readers. Of course, a fraction of transistors in the CMOS processor is sleeping at any given moment in order to prevent overheating. This was not taken into account in our estimate. The same approach can also be implemented in SFQ electronics.

| Unit | $f_{cl}$ (GHz) | $N$ ($10^9$ JJs or transistors) | $P_{cold}$ at 4.2 K (W) | $P_{hot}$ (W) lower bound $\eta = 0.2$ | $P_{hot}$ (W) upper bound $\eta = 0.02$ |
|---|---|---|---|---|---|
| SFQ | 4 | 1.4 | 0.71 | 250 | 2500 |
| i7-4790 | 4 | 1.4 | - | 88 | 88 |

Because this result may look too pessimistic, we decided to check it by comparing the power requirements per GFLOPs in CMOS and SFQ implementations. The only existing data for operational bit-serial, single-precision (32-bit) floating-point adders (FPA) and floating-point multipliers (FPM) made in RSFQ technology were reported in [69]. They have, respectively, 16830 JJs and 18766 JJs. (For a comparison, a 32-bit adder requires about 3000 transistors.) The measured performance is shown in Table IV along with the data for an i7-4790 Haswell CPU. We used the above cited thermal efficiencies of the cryocooler to get the lower and upper bounds of power consumption and computation efficiency (in GFLOP per joule) at room temperature. Basically, we get the same result as above: RSFQ-based FPA and FMP are 2 to 20 times less efficient than the off-shelf CPU, Table IV. The numbers reported in [69] correspond to the RSFQ circuits using bias resistors with significant static power dissipation. If this dissipation is eliminated by using, e.g., ERSFQ approach, the computation efficiency may improve by a factor of ~10, making the SFQ circuits competitive if implemented in a very-large-scale system, but still losing efficiency if used in a small-scale system.

| Unit | $f_{cl}$ (GHz) | Through-put (GFLOPS) | Power at 4.2 K (mW) | Power dissipation at room-$T$ (W) | Efficiency at room-$T$ (GFLOP/J) |
|---|---|---|---|---|---|
| FPA | 58 | 2.23 | 4.92 | 1.7 to 17[a] | 0.13 to 1.3 |
| FPM | 59 | 2.36 | 5.76 | 2.0 to 20[a] | 0.12 to 1.2 |
| i7-4790 | 4 | 200 | - | 88 | 2.27 |

[a] Using the inverse efficiencies of 352 ($\eta = 0.2$) and 3520 ($\eta = 0.02$).

If the average critical current in SFQ circuits can be reduced by a factor of five, to 35 µA, somehow keeping the acceptable bit error rate determined by the smallest junctions in the circuit, the total energy dissipation in complex SFQ circuits with account for refrigeration can become somewhat lower than in their CMOS counterparts. Note, however, that we have not accounted for any power consumption associated with auxiliary electronics, thermal radiation, and heat conduction via cryogenic cables.

A more optimistic estimate of computation efficiency of SFQ processors can be obtained by estimating naively the energy consumption per PFLOP, using the energy per single-bit operation $<E_{SBOP}>$ estimated above. Depending on the circuit architecture, a 32-bit floating-point operation requires about 1500 single-bit operations (for adders) and somewhat more for multipliers. Then, the lower bound on the energy loss per FLOP at 4.2 K is $E_{FLOP} \approx 2000 < E_{SBOP} > \approx 5 \cdot 10^{-15}$ J, or 5 J per PFLOP, giving a computation efficiency of ~ 0.2 PFLOP/J at 4.2 K or from about 57 to 570 GFLOP/J at 300 K. This is from 25 to 250 times better than CMOS, see Table IV. Unfortunately, there are currently no architectural solutions to realize this ultimate computation efficiency because superconductor electronics does not have compact and efficient memory, whereas memory is abundant in semiconductor computers.

So, it follows from the estimates above that cryogenic computational systems based on the SFQ logic versions utilizing JJ switching and emulating CMOS architectures will likely be faster than the CMOS-based but not necessarily more energy efficient. This has a very simple reason - requirements for switching elements used in any computing system are basically the same and independent of the operating temperature. These are requirements of reliability, drivability, and communications: fast switching times ~ 1 ps, immunity to the thermal noise and parameter spreads (low bit error rates), and ability to drive other parts of the system and communicate with them. These requirements set the minimum ratio $E_{sw}/k_B T$, typically ~ $10^4$, for the switches used in a computer, where $T$ is its operating temperature. Then, of two computers operating with the same $E_{sw}/k_B T$ ratio and having similar architectures, the one requiring refrigeration to $T_{cold}$ may become more energy efficient than a computer operating at $T_{hot}$ only if the refrigerator is nearly ideal, $\eta \geq 1 - T_{cold}/T_{hot}$, which is not possible at cryogenic temperatures.

### B. Adiabatic Quantum Flux Parametron (AQFP) circuits

Among the superconducting digital technologies developed to date only AQFP circuits can be truly energy efficient. AQFP is the same QFP invented by E. Goto more than 30 years ago, see [83]-[84] and references therein, but operated in a slow (adiabatic) regime [85],[86]. Instead of using JJ switching to move fluxons, as in RSFQ and RQL, the information in QFP is encoded by a fluxon location in a double-well potential, in the left ("0") or the right ("1") well, and moved by adiabatically varying the shape of the potential, using ac bias currents. The measured energy dissipation at 5 GHz operation is extremely low, ~ $0.1 I_c \Phi_0$ per bit [147] and can be further reduced. QFP is very similar in the operation principle to a much older device – parametric quantron [148],[149]. Theoretically, these types of parametric devices can provide for the lowest energy consumption in computations [150].

Because parallel pipelining is very natural for AQFP, processors with higher computational efficiency than ERSFQ, RSFQ, and RQL can be designed [151]. According to estimates in [151], the computational energy efficiency of some algorithms implemented in AQFP could be 7 orders of magnitude higher, with account for refrigeration, than of CMOS circuits.

## C. Summary

A very inquisitive reader may ask why our assessment of energy efficiency of RSFQ and RQL technologies for high-performance computing differs from the one made in [16], which indicated that RQL might be able to meet the efficiency requirement. The difference comes from simple arithmetic. In the simplistic estimate of the power required to compute 1 PFLOPS, see Eq. (1) in [16], the junction activity factor $\alpha$ was double-counted: the first time implicitly in the estimate of energy per switch in the RQL $(1/3)\langle I_c\rangle\Phi_0$, which already includes $\alpha=1/2$; and the second time explicitly in the power estimate based on the number of gates per FLOPS. Also, in this estimate, the minimum critical current of JJs of 25 µA was used instead of the weighted average $\langle I_c\rangle$ entering (16)-(19), which is typically a factor of two larger than the minimum $I_c$. As a result the power per PFLOPS was probably underestimated by about a factor of four, and the energy efficiency was overestimated by the same amount. The use of the minimum $I_c$ in energy-consumption estimates is very typical for reviews of SCE; see for example [77],[79],[118]. It implies that a superconducting computer (circuit) can be built from the identical junctions having the minimum critical current and the minimum area. This could suffice for an order-of-magnitude estimate, but it is not a valid assumption.

## V. FUTURE WORK

The choice of the most promising directions of future work strongly depends on the strategic goals one wants to accomplish. In a limited funding environment, setting the priorities and optimizing the strategy should be done thoroughly because "no one can have his cake and eat it too," and diverting time, efforts, and funding toward one area may leave the other one starving, and so on. The author's selection is given below and should not be construed in any form as funding recommendations.

### A. Energy-Efficient Computing

If the primary goal is energy efficiency, then a clear winner among the existing and relatively mature technologies is AQFP, because its energy consumption, including refrigeration can be made a few orders of magnitude lower than in the existing room-temperature technologies. However, the clock speed of truly energy-efficient AQFP circuits is limited to ~ 7 GHz. Due to the use of multiple transformers and ac power, the cell area is currently large and the integration density is low. The limits on the integration scale are the same as for other types of SCE as described in the previous sections. Unfortunately, there is currently no work on AQFP circuits for high-performance computing anywhere except Japan. Interestingly, however, most of the control circuitry in D-Wave quantum annealing processors operating at tens of mK is based on QFPs due to their ultralow power dissipation.

Even higher energy efficiency is promised by reversible (or almost reversible) computing. Ideas of reversible computing with superconducting circuits were discussed in [107]-[110]. Simple circuits, shift registers, with energy-per-bit near the Landauer's thermodynamic limit $k_BT\ln2$ have already been demonstrated [110]. Larger circuits with richer functionalities need to be developed and fabricated. Unfortunately, the progress in the area of superconducting reversible computing has been slow due to lack of funding.

As was shown above, the energy efficiency of SFQ processors utilizing junction switching and CMOS architectures, i.e., designed by replacing CMOS logic cells by SFQ logic cells, is marginal, and the energy saving may not be sufficient to warrant the effort. Therefore, the obvious way of making SFQ processors more energy efficient is to employ different information processing solutions and architectures that would require significantly fewer JJs than transistors to implement, and would not require external memories. These ideas have been discussed to some extent but require practical development. For instance, V.K. Semenov in [105] argued that RSFQ blocks should be implemented in a way that preserves their inherent logic and memory functions, and makes use of the simplicity and record-high speed of SFQ T-flip-flops [106] instead of replicating CMOS logic cells. Interesting to note, the RSFQ technology inventors stated in their original comprehensive review [79] that "a universal von-Neumann-type computer is probably the worst device for implementation using the RSFQ (or any other superfast) technology," and explain why. Somehow this message and the idea that RSFQ blocks with their logic/memory functions are cellular automata or finite-state machines rather than CMOS-type logic gates were forgotten in the course of the last 25 years.

Unfortunately, the only idea that is being actively worked on is the most trivial one – reducing the average critical current $\langle I_c\rangle$ in SFQ cells. This reduction has a clear limit ~ 50 µA, below which the circuit density, bit error rate, and circuit yield will be substantially compromised, and hence cannot be a long-term strategy. On this road, RQL has currently a big advantage because of the serial biasing. For instance, SFQ circuits with the largest JJ count per chip demonstrated so far have been the RQL shift registers [76] made by the MIT-LL SFQ4ee process [28]. It does not mean that RQL is a better technology however, only that its problems are in a different area. RSFQ-based circuits are clearly behind in JJ count due to the parallel biasing and associated problems. Therefore, the prime goals should be in developing serial-biasing (current-recycling) solutions suitable for multimillion-JJ circuits. Without solving this problem, RSFQ-based circuits, in the author's opinion, have no future in high-performance computing or other applications requiring VLSI.

### B. High-Speed Computing

On the other hand, if the primary goal is the computation speed, high clock frequency, and energy dissipation is secondary, then the technology selection and the development priorities are very different. RSFQ is clearly the fastest digital technology developed so far, and capable of reaching ~ 70 GHz clock frequencies in the current technology node and over 100 GHz if the $J_c$ is increased to 0.5 mA/µm² and beyond. Therefore, development of current recycling for VLSI circuits becomes the priority number one. Since resistive biasing has no place in superconductor VLSI due to heating, ERSFQ becomes a clear winner if its inductive biasing approach can be scaled up to the VLSI. RQL and AQFP circuits will probably lose the speed competition because of the multiphase ac biasing.

Design development priorities in this case are also different. Instead of reducing the critical current $<I_c>$, it should be increased and optimized for junctions with higher $J_c$, increased to ~ 0.5 mA/$\mu$m$^2$ and above, perhaps self-shunted JJs, which will likely have larger parameter spreads at the same sizes as the current tunnel junctions with $J_c = 100$ $\mu$A/$\mu$m$^2$.

As priority number two, I would rate the development of new architectural solutions that do not copy the standard CMOS. It is the author's opinion that SCE electronics cannot win or even be competitive if it mimics CMOS, because of the five orders of magnitude difference in integration scale.

Number three on my list would be the development of fast and compact JJ-based memories that do not use magnetic materials and magnetic tunnel junctions discussed in [111]-[113], a development that may take many years. Josephson random access memories (RAM) have been demonstrated in older process nodes with large features and low RAM densities [23], [114],[115], [152]. They should be improved and implemented in the advanced process nodes using more advanced materials and smaller features.

## C. VLSI Technology Development

The main advantage of RSFQ-based processors is that they can run much faster (likely 25 times) than the 4 GHz offered by CMOS because energy dissipation on the chip is significantly reduced, see (17)-(18), and moved instead to a much larger cryocooling system at room temperature. So, our hypothetical chip with 1.4B JJs can run at 100 GHz and dissipate only about 18 W at 4.2 K. The typical power that can be removed from the chip without raising its temperature more than 1 K in liquid He is ~ 1 W/cm$^2$. So this chip can be cooled if its active area is larger than ~ 18 cm$^2$. This should be an easy task because a SCE chip with 1.4B RSJs would have an area of 700 cm$^2$ at the present maximum density of 2M RSJs per cm$^2$. A chip of this area is difficult to imagine and would be impossible to manufacture, so a superconducting multichip module (MCM) with about 200 of 2-cm$^2$ chips could be dreamed of as an equivalent. The required MCM technology with SFQ interchip communication data rates exceeding 100 GHz has been demonstrated; see [116],[117] and references therein.

This example demonstrates that the development of VLSI technology for SCE trumps all the priorities mentioned above. Without solving the scalability problem, development of SFQ processors has no merit. Note that fabrication technology development and implementation of new materials and processes are usually much more expensive and time consuming than circuit design because the former requires expensive and sophisticated processing equipment whereas the latter requires only good ideas and engineers with CMOS-based computers. This simple truth was mainly ignored during the last 25 years.

Many ideas of what could be done have been floated around. I briefly review them in order to rate their impact on increasing the circuit density and feasibility of implementation. To be specific, let us set an increase in the circuit density by a factor of ten, i.e., achieving $2 \cdot 10^7$ cm$^{-2}$ density of JJs and inductors, as the primary near-term goal. (The author simply cannot imagine a 200-chip MCM for our 1.4B-JJ processor, but can imagine a 20-chip MCM.)

### Self-shunted, high-$J_c$ junctions

Getting rid of resistive shunts would have the largest impact on the circuit density, as is clear from Fig. 7, and would shorten the fabrication process by eliminating the resistor module, Fig. 5. This requires replacing the hysteretic *SIS* tunnel junctions with nonhysteretic junctions. In this context, we would like to clear up one of the misconceptions in this area, which appeared in [77], [119],[153] and crept into many other publications. It is an incorrect assertion that all tunnel junctions become self-shunted at high $J_c$ , e.g., ~ 100 kA/cm$^2$ in the case of Nb-based junctions. This misconception is a result of a simple-minded application of the McCumber-Stewart [89],[90] parameter $\beta_c = 2\pi I_c R_n^2 C/\Phi_0$, derived for a linear shunt resistor, to tunnel junctions with nonlinear, voltage-dependent, damping (2). Then, using $R_N$ as a damping resistance at all voltages, noting that the expression can be rewritten as $\beta_c = 2\pi (I_c R_N)^2 C_s/(J_c \Phi_0)$, where $C_s$ is the junction specific capacitance, and that $I_c R_N$ is the electrode material property ~ $\Delta/e$, it may appear that $\beta_c \leq 1$ can be obtained by mere increasing the $J_c$. This is, of course, incorrect because there are not enough electronic states at $V < V_g$ to provide damping ($R_{sg} >> R_N$), and switching back into $V=0$ state is hysteretic. Therefore, self-shunting can be induced only by increasing the density of states in the gap or in the barrier allowing for a substantial ohmic conduction (junction "leakage") at $V < V_g$. This can be achieved by introducing defects in the tunnel barrier, e.g., pin-holes and oxygen vacancies [120]-[124], or by using junction barriers with direct conduction such as normal metals, doped semiconductors, etc. However, high-quality tunnel junctions, e.g., using Nb$_2$O$_5$ and AlN$_x$ barriers instead of AlO$_x$, remain highly hysteretic at $J_c$ values much larger than 100 kA/cm$^2$ [125]-[127]. This makes Nb/Al-AlN$_x$/Nb tunnel junctions unattractive for VLSI applications requiring self-shunted JJs, although they are perfectly good junctions for *SIS* detector applications.

Among junctions with direct conduction, relatively high values of $V_c$ required for digital applications have been demonstrated by Nb-based junctions with amorphous Si barriers doped by various impurities creating deep levels near the middle of the band gap, like Nb, W, etc.; see [128] and references therein, [129-[131], and many older works, e.g., [132]-[134]. From the author's point of view, their only advantage is that, at large levels of doping (~ 10%), self-shunting and nonhysteretic behavior is obtained. The other properties are rather drawbacks. Their $V_c$s are inferior to Nb/AlO$_x$/Nb junctions at the same $J_c$, and the temperature dependence of the critical current is stronger than in *SIS* junctions. The self-shunting in these devices is likely due to inelastic and resonance tunneling via multiple localized sates, percolation paths, within the barrier [128],[135]. As a result of this unusual conduction mechanism, the *I-V* characteristics are nonlinear and deviate from the RCSJ model; the devices also have unusually high specific capacitance, which is larger than in $SIS$ devices. The circuit clock speeds based on these devices are also reduced by ~ 30% with respect to the same circuits made with $SIS$ junctions [130]. It is not clear whether α-Si-doped devices are reproducible at submicron dimensions required for their implementation in VLSI circuits, because of the percolation-type current transport and possible fluctuations in the number of localized centers and resonant paths. A good uniformity of α-Si:Nb barriers in voltage standard applications was demonstrated using JJs with areas of hundreds of square micrometers. These data cannot be projected onto submicron-scale devices.

Therefore, the most important task is to evaluate the critical current spreads of α-Si-doped JJs with submicron dimensions at critical current densities in the range from 0.1 mA/μm² to about 1 mA/μm² by using the modern processing tools similar to the one used in [28-30]. Somehow this has not been done despite these junctions being around for over 35 years. (Sperry Corporation was trying to commercialize superconducting devices and circuits using Nb and NbN junctions with doped α-Si, α-Ge, and Si-Ge barriers in the 1980s.) The same comment applies to high-$J_c$ Nb/AlO$_x$-Al/Nb junctions. Their subgap transport is due to multiple Andreev reflections via defects, most likely oxygen vacancies, in the barrier. So they could be viewed as approaching the metal-insulator transition from the opposite side than α-Si-doped barriers. There has been a claim that, at high-$J_c$, the transport mechanism in Nb/AlO$_x$-Al/Nb junctions becomes universal [136], so the junction-to-junction variations could be small. However, the experimental basis for this is about 200 high-$J_c$ junctions in [92],[137], which is not sufficient to build a VLSI technology on. So, the near-term goals should be the evaluation of junctions with $J_c = 0.5$ mA/μm², five times larger than it is in the SFQ5ee process, in order to measure their parameter spreads and self-shunting.

One of the drawbacks of all compact self-shunted JJs is self-heating. Heating in the RSJs is negligible because the $E_{sw}$ is dumped into the resistor whose area is much larger than the JJ area. In the self-shunted JJs the same energy dissipates inside the junction and creates nonequilibrium quasiparticles. Energy density and heating increase with increasing $J_c$. The maximum clock frequency of these devices may then be determined by the energy relaxation rate in order to prevent memory effects and time jitter associated with the influence of one switching on the next and difference in activity factors in different parts of the circuit.

### Number of junction layers

Increasing the number of independent JJ layers to two is beneficial and may increase the JJ density by a factor of 2x. The work in this direction is currently underway at AIST in Japan. A larger increase would likely have no proportional effect on the density because of the need for interconnecting and through vias reducing the available area.

### Compact inductors

The next in importance is the increase in inductor density $n_L$. This can be done by decreasing their area, increasing the inductance per unit length, and increasing the number of layers, $m_L$. Increasing $m_L$ above two will have little impact because of the vias and cross-talk problems. At small spacing between the inductors ~ 0.25 μm, the cross-talk between the inductors on the same layer or between the layers becomes significant. So the inductors must be isolated by the ground planes and vias between them, boxed into a coaxial-type configuration. This reduces the maximum density and defeats the purpose.

Let us take as an example, the average inductor in the cells of the AC-biased shift register in [75],[99] and shown in Fig. 12. Its value is about 6 pH, and in the current process it occupies the area of ~ 8.5 μm². A factor of ten increase in $n_L$ means that its area needs to be reduced to ~ 0.85 μm². This can be achieved by using a thin layer of a high-kinetic-inductance material like MoN$_x$ or NbN$_x$ for signal inductors, similar to our SFQ5ee process for bias inductors [30]. It is well known that the kinetic inductance of thin superconducting films $L_k = \mu_0 \lambda^2/t$ is much larger that their geometric inductance if $\lambda >> t$ and $\lambda^2/t >> w$, where $\lambda$ is magnetic field penetration depth, $t$ and $w$ are the film thickness and width, respectively. The kinetic-inductor layer can be placed in a close proximity to JJs, e.g., between layers J5 and M6 instead of the layer of resistors, R5 in Fig. 2, since with self-shunted JJs resistive shunts will no longer be required. The $I_c$ of this kinetic inductor should be a factor of 2x larger than $I_c$ of JJs, or about 0.25 mA. Then, using a typical value of inductance per square of ~ 5 pH/sq, for a 60-nm-thick MoN$_x$ film, and the film width of 0.5 μm to guarantee the critical current, the kinetic inductor length will be 0.6 μm and the area, accounting for a 0.25-μm line spacing, will be about 0.5 μm². Adding vias to the inductor will double the area. A compact via process has been demonstrated [138]. So one layer of kinetic inductors near the self-shunted JJs can increase the circuit density by a factor of ten, and two layers by a factor of ~ 20x.

The use of kinetic inductors does not help in reducing the area of inductive couplers and transformers in RQL and AQFP circuits, which depend on the geometric mutual inductance. The author has no readily available solutions for increasing the density of RQL and AQFP circuits by a factor of ten, except for reducing the minimum linewidth and spacing of inductors down to ~ 100 nm.

### JJs as inductors

Nonswitching Josephson junctions can be used as circuit inductors. This idea is very old, but so far has been only implemented in superconducting persistent-current qubits [139]-[140] and their advanced versions [141], and in tunable RF filters [142]. A segment of Josephson transmission line with JJs replacing all inductors would look like the one shown in Fig. 13.

The area saving can only be achieved if a vertical stack of inductor junctions is used. In order to preserve the design of all main SFQ cells, the minimum number of junctions in the stack should be three. Then, a quantizing inductor with a $2\pi$ total phase shift can be formed using two $L_J$ units shown in Fig. 13 (top panel). The phase across each JJ in the stack will be $\sim\pi/3$, far enough from $\pi/2$ when the critical current is reached. We would like to term a switching JJ and three stacked nonswitching inductor-JJs a complementary pair, or a *CLJJ* pair, as shown in Fig. 13. Accordingly, circuits built using this technology can be termed complementary-SFQ circuits or CSFQ circuits.

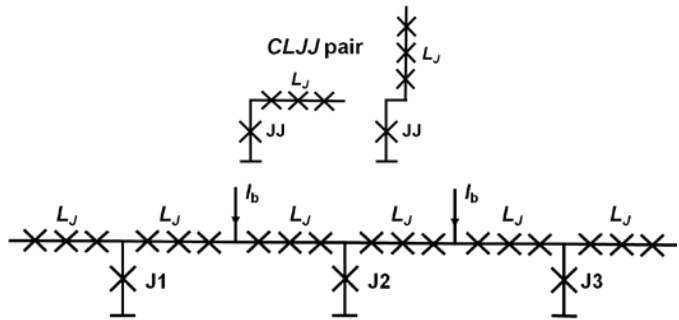

Fig. 13. A complementary JJ-inductor pair, a *CLJJ* pair, formed by a junction and a vertical stack of three nonswitching JJs, $L_J$. The ability to apply bias and signals at the point between the JJ and $L_J$ is important and shown explicitly in the circuit diagrams in the top panel. Bottom panel: a segment of Josephson transmission line (JTL) using *CLJJ* pairs, i.e., with all inductors replaced by series arrays of nonswitching Josephson junctions $L_J$. The switching junctions are J1, J2, etc., $I_b$ is the bias current.

Since there is no switching speed requirement for the inductor-junctions in the stack, the $V_c$ of the junctions can be lower than of the switching JJs. Then, a much simpler technology of *SNS* junctions can be used to form the stack, where *N* is a normal metal compatible with the process, e.g., Al, Mo, Ti, etc. No doped α-Si or other fancy barriers are required. They could also be used, but are not necessary.

The ability to apply bias and signal at the point between the JJ and $L_J$ is important and shown explicitly in the circuit diagrams. The fabrication process can be sketched briefly as shown in Fig. 14.

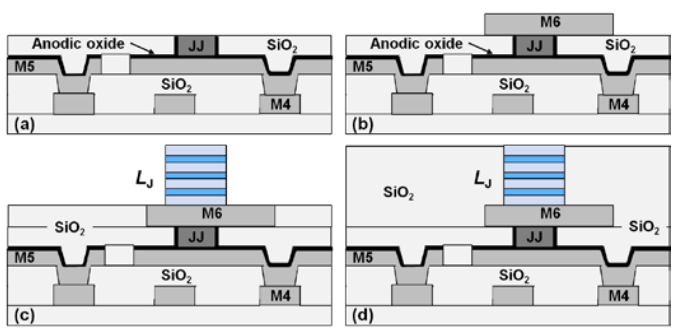

Fig. 14. A process for making complementary $L_J$-JJ pairs, *CLJJ* pairs: regular JJs coupled to inductors formed by three stacked nonswitching JJs. After finishing the switching JJ planarization step, see Fig. 4(h), the resistor module in Fig. 5 can be abandoned because we plan to use self-shunted JJs. Then, a wiring layer M6 will be deposited and patterned (b). After M6 planarization by the dielectric CMP, a multilayer Nb-M-Nb-M-Nb-M-Nb will be deposited on the planar surface and patterned to form a vertical stack of three JJs, $L_J$, where M is the barrier metal (c). Finally all $L_J$ stacks will be planarized to their tops, similar to the JJ planarization described in the text. Top wiring will be deposited and patterned to interconnect the JJ-inductors, not shown here.

The typical critical current of nonswitching JJs, $I_{cL}$ can be estimated from the typical value of $\beta_L \sim 2$, using instead of magnetic inductance *L* the Josephson (kinetic) inductance $L_J = 3\Phi_0/(2\pi I_{cL})$, which gives $I_{cL} = (3/\beta_L)$, or $I_{cL} \sim 1.5 I_c$. To satisfy our 10x density increase requirement, the $L_J$ area should be less than 1 μm², so the critical current density of nonswitching JJs should be larger than 100 μA/μm².

The process shown in Fig. 14 would be relatively easy to implement. However, the drawback of replacing thin-film inductors with JJ-inductors is that a very simple and highly reliable device – a narrow strip of high-kinetic-inductance material – is replaced by much more complex and less reliable devices using a multilayered stack of JJs. The JJs in the stack need to have (nearly) identical critical currents. This is a serious complication of the process and a potential impediment to the process reliability and yield. Moreover, implementation of *CLJJ* pairs does not solve the problem of miniaturization of inductive couplers and transformers in RQL and AQFP circuits.

### *Phase shifters and pi-junctions*

The use of phase shifters based on pi-junctions with ferromagnetic barriers has been discussed and demonstrated in [143]-[145]. Despite their useful features and interesting physics, adding pi-junctions would have a negative impact on the circuit density, because the critical current density of pi-junctions is a factor of ~100 lower than in regular 0-junctions and their area accordingly is a 100x times larger. The existing advanced fabrication processes have so many superconducting layers that this function (phase shifting) could be easily realized by using a miniature rings and narrow wires to provide magnetic bias; see, e.g., [146].

## VI. CONCLUSION

In order to evaluate whether SFQ electronics is scalable to VLSI levels required for achieving computation complexities comparable to CMOS processors, we have reviewed the current state of the fabrication technology for superconducting digital circuits based on single-flux-quantum encoding of information. We have described the limitations imposed on the circuit density by Josephson junctions, circuit inductors, shunt and bias resistors, parameter spreads, etc. We have shown that the currently achievable maximum circuit density of resistively shunted Josephson junctions and inductors is about $2 \cdot 10^6$ cm⁻², which is almost four orders of magnitude lower than the density of transistors in modern CMOS circuits. We have described the fabrication-process development required for increasing the density of SFQ digital circuits by a factor of ten, including self-shunted Josephson junctions, kinetic inductors, complementary JJ-inductor pairs (*CLJJ*) using Josephson inductance of stacked, nonswitching junctions, etc. Energy dissipation in superconducting circuits has been also reviewed in order to estimate whether SFQ electronics, which requires cryogenic refrigeration, can be energy efficient in comparison with CMOS. We estimated that energy dissipation in SFQ circuits based on JJ switching, such as RSFQ, ERSFQ, RQL, and having comparable complexity to the modern CMOS processors will be comparable to that in CMOS processors when refrigeration energy is taken into account. The energy efficiency can be improved if innovative circuit architectures could be developed, which use much fewer JJs than transistors for the same information processing and have minimum use of external memories. The most energy efficient technology is adiabatic quantum-flux parametron (AQFP), which can run at ~ 7 GHz clock frequency. RSFQ and its "energy efficient" versions remain the fastest digital superconductor technology capable of clock frequencies over 100 GHz, if energy efficiency is not a primary goal. However, it cannot be scaled to VLSI until serial-biasing and current-recycling VLSI technology are developed. The main scalability problem of SFQ digital electronics stems from its advantages – magnetic flux information encoding and SFQ voltage pulse transferring, resulting in large-area SFQ cells. It is clear that confining magnetic flux takes much larger volume and effort than confining charge in the gates of CMOS transistors. As a result, the complexity of SCE digital electronics is expected to remain relatively low unless unforeseen breakthroughs happen. Superconductor electronics has many advantages in applications where the cryogenic environment is mandatory and dictated by the system performance requirements that cannot be met by any other means. For instance: as control electronics and cryogenic data processors for very large arrays of cryogenic sensors; control electronics for analog quantum computing based on superconducting quantum annealers; for gate-based quantum computing with superconducting qubits; for application-specific ultrafast circuits. There is no doubt that (nearly) reversible superconducting circuits, approaching and crossing the thermodynamic limit, will soon be demonstrated as well as many other interesting circuits. However, the impact of SFQ electronics on general-purpose and high-performance computing, in my opinion, will remain low in the foreseeable future because of the insufficient scale of integration.

## ACKNOWLEDGMENT

I am very grateful to all my colleagues at MIT Lincoln Laboratory who are involved with fabrication process development for superconductor electronics, especially to Vladimir Bolkhovsky and Scott Zarr, to Terry Weir and Alex Wynn for their part in device testing, and to Mark Gouker and Leonard Johnson for the discussions and management of the program. I would like to thank Vasili K. Semenov, Alex F. Kirichenko, Timur Filippov, Quentin Herr, Marc Manheimer, and D. Scott Holmes for useful discussions. My special thanks are to Daniel E. Oates for reading the manuscript and suggesting numerous improvements.

This research is based upon work supported by the Office of the Director of National Intelligence (ODNI), Intelligence Advanced Research Projects Activity (IARPA), via Air Force Contract FA872105C0002. The views and conclusions contained herein are those of the author and should not be interpreted as necessarily representing the official policies or endorsements, either expressed or implied, of the ODNI, IARPA, or the U.S. Government. The U.S. Government is authorized to reproduce and distribute reprints for Governmental purposes notwithstanding any copyright annotation thereon.